# Photocorrosion-limited maximum efficiency of solar photoelectrochemical water splitting

Ling-Ju Guo[1], Jun-Wei Luo*[2,3], Tao He[1], Su-Huai Wei[4], and Shu-Shen Li[2,3]

*[1]Chinese Academy of Sciences (CAS) Key Laboratory of Nanosystem and Hierarchy Fabrication, CAS Center for Excellence in Nanoscience, National Center for Nanoscience and Technology, Beijing 100190, China*

*[2]State key laboratory of superlattices and microstructures, Institute of Semiconductors, Chinese Academy of Sciences, Beijing 100083, China*

*[3] College of Materials Science and Opto-Electronic Technology, University of Chinese Academy of Sciences, Beijing 100049, China*

*[4]Beijing Computational Science Research Center, Beijing 100094, China*

*Email: jwluo@semi.ac.cn

## Abstract

**Photoelectrochemical (PEC) water splitting to generate hydrogen is one of the most studied methods for converting solar energy into clean fuel because of its simplicity and potentially low cost. Despite over 40 years of intensive research, PEC water splitting remains in its early stages with stable efficiencies far less than 10%, a benchmark for commercial applications. Here, we revealed that the desired photocorrosion stability sets a limit of 2.48 eV (relative to the normal hydrogen electrode (NHE)) for the highest possible potential of the valence band (VB) edge of a photocorrosion-resistant semiconducting photocatalyst. We further demonstrated that such limitation has a deep root in underlying physics after deducing the relation between energy position of the valence band edge and free-energy for a semiconductor. The disparity between the stability-limited VB potential at 2.48 V and the oxygen evolution reaction (OER) potential at 1.23 V *vs* NHE reduces the maximum STH conversion efficiency to approximately 8% for long-term stable single-bandgap PEC water splitting cells. Based on this understanding, we suggest that the most promising strategy to overcome this 8% efficiency limit is to decouple the requirements of efficient light harvesting and chemical stability by protecting the active semiconductor photocatalyst surface with a photocorrosion-resistant oxide coating layer.**

## I. Introduction

Photoelectrochemical (PEC) cells that do not use an external bias or sacrificial reagents are of interest because of their simplicity and potentially low fabrication costs[1-4]. A PEC cell is based on a semiconductor/liquid junction, where free carriers (electrons and holes) generated by light absorption in the semiconductor are driven into the solution by the electric field at the junctions. Specifically, for a n-type semiconductor photocatalyst PEC cell, photogenerated holes initiate the oxygen evolution reaction (OER) at the surface of the semiconductor electrode and oxidize water to oxygen; photogenerated electrons entering the counter electrode drive the hydrogen evolution reaction (HER) and reduce water to hydrogen. A semiconducting material that can efficiently absorb sunlight to generate electron-hole pairs and that has a high mobility and long lifetime to prevent electron-hole recombination within the material before the electron-hole pairs reach the junction is needed to achieve efficient solar-to-hydrogen (STH) conversion. Additionally, the bandgap of a semiconductor photocatalyst must straddle the HER and OER potentials[1] to afford favorable reaction kinetics toward overall water splitting. In other words, the potential of the conduction band (CB) edge ($\phi^{\mathrm{CB}}$) must be more negative than the HER potential ($\phi^0(\mathrm{H^+/H_2}) = 0\ \mathrm{V}$ *vs* NHE, which is approximately 4.5 eV below the vacuum level; NHE = normal hydrogen electrode), and the potential of the valence band (VB) edge ($\phi^{\mathrm{VB}}$) must be more positive than the OER potential ($\phi^0(\mathrm{H_2O/O_2}) = 1.23\ \mathrm{V}$ *vs* NHE). In addition to exhibiting an optimal bandgap for harnessing sufficient sunlight and favorable band-edge positions for driving overall water splitting, a semiconductor photocatalyst must also exhibit excellent stability in aqueous solution, which is the main limiting factor for the use of conventional photovoltaic semiconductors as photocatalysts[1-5] and thus is the main challenge to realizing high STH efficiency for water splitting.

Silicon, CdTe, and III-V semiconductors are important materials in the photovoltaic industries because they are excellent light absorbers with high carrier mobilities, but their stabilities in aqueous solution are limited (<days)[5-7]. On the other hand, wide-bandgap oxides, such as $TiO_2$, $SrTiO_3$, and $WO_3$, are highly resistant to photocorrosion in aqueous solution, but they suffer from poor PEC performances because their bandgaps are too wide to absorb a sufficient amount of the solar spectrum[1-5]. Thus far, no material capable of harvesting the whole visible-light spectrum with stability against photocorrosion has been found. Current knowledge is insufficient to identify semiconductor photocatalysts simultaneously possessing efficiency and stability. Thus, empirical approaches are often invoked with a hope to discover effective photocatalysts, such as using high-throughput experimental[3] or computational[8, 9] screening of numerous materials as solar water-splitting photocatalysts. In the end, hundreds of thousands of semiconducting compounds will likely be examined and tested for PEC efficiency and stability without the certainty of success. Chen and Wang[10] recently examined the stability of thirty photocatalytic semiconductors in aqueous solution using the method developed by H. Gerischer[11] in 1977 in combination with the first-principles calculations and discussed their thermodynamic stabilities and trends against the oxidative and reductive photocorrosion. The primary aim of this study is to address whether a semiconductor photocatalyst can be simultaneously efficient and stable and thereby shed light on the knowledge-based design of highly efficient PEC water-splitting cells.

Many of the most-stable metal-oxide PEC photocatalysts possess wide bandgaps because the width of the bandgap is argued to be a measure of the chemical bond strength[1]. However, GaN and ZnS have even larger bandgaps than stable $TiO_2$ and $WO_3$, but these semiconductors exhibit photodecomposition in aqueous solution[1, 2, 4, 5]. The high-lying VB energy of both GaN and ZnS, where the VB edge consists of high-energy N 2p or S 2p orbitals, is a prominent feature distinguishing these semiconductors from $TiO_2$ and $WO_3$, where the VB edge is dominated by low-energy O 2p orbitals and stays low and somewhat nails at approximately 3.0 V *vs* NHE. $Cu_2O$ is a rare metal oxide providing favorable energy positions and an optimal bandgap for PEC water splitting resulting from a high-lying VB, which is composed of primarily Cu 3d orbitals instead of O 2p orbitals, but $Cu_2O$ is also unstable in aqueous solution[12]. Thus, we speculate that the energy position of the VB edge, rather than the width of the bandgap, determines the stability of a semiconductor under illumination in aqueous solution: a more positive VB potential (lower energy) results in better thermodynamic stability. Because the CB potential must be negative *vs* NHE, a more positive VB potential indicates a wider bandgap and hence less efficient PEC photocatalysis owing to less harvested sunlight. Therefore, a proper balance must be attained between the stability and efficiency of a semiconductor for PEC water splitting. To reveal the hidden relationship between the VB potential and the thermodynamic stability for semiconducting PEC photocatalysts, here, we have examined 202 semiconductors that are known to either catalyze overall water splitting or reduce or oxidize water in the presence of an external sacrificial redox agent as collected in three most comprehensive, to our best knowledge, review articles [13-15]. Note that we don't have any bias in choosing the known catalytic materials in our study. We found that the desired photocorrosion stability sets a limit of the highest possible VB potential at 2.48 V *vs* NHE for semiconducting PEC photocatalysts, which in turn excludes PEC photocatalytic materials with bandgaps less than 2.48 eV. We also demonstrated that such limitation has a deep root in underlying physics after deducing the relation between energy position of the valence band edge and free-energy for a semiconductor.

## II. Methods

**Assessing the thermodynamic stabilities of semiconductors against photocorrosion.** The thermodynamic stability of a semiconductor in contact with an electrolyte solution is characterized by the decomposition redox potential, which is the required minimum Fermi energy of the electrons and holes driving the reductive and oxidative decomposition reactions, respectively[10, 11]. By analyzing the relative positions of competing potentials of water redox reactions, semiconductor decomposition reactions, and semiconductor band edges, we can determine the stabilities of the semiconductors in the PEC process[10, 11]. For an optimized semiconductor exhibiting a bandgap straddling both HER and OER potentials, Figure 1a illustrates the thermodynamic stability requirement that the reductive decomposition potential $\phi^{re}$ of a semiconductor must be more negative than the HER potential $\phi^0(\mathrm{H}^+/\mathrm{H}_2)$ for water reduction, where photogenerated electrons prefer to participate in water reduction to produce $H_2$, and that the oxidative decomposition potential $\phi^{ox}$ of a semiconductor must be more positive than the OER potential $\phi^0(\mathrm{H_2O/O_2})$ for water oxidation, where photogenerated holes prefer to participate in water oxidation to produce oxygen. [Note that Figure 1a shows a case of semiconductor $\phi^{ox}$ being above its VB edge but below the $\phi^0(\mathrm{H_2O/O_2})$, which was assigned by Gerischer as metastable protected by solvent in his original paper[11]. This different assignment may lead to overestimation of the stability of semiconductor photocatalysts but will not affect our conclusions as we will discuss below.] Otherwise, as shown in Figure 1b-d, photogenerated electrons or holes prefer to drive the self-decomposition of the semiconductor instead of the water redox reactions, resulting in instability under illumination in aqueous solution. In this case, a semiconductor has at least one oxidative decomposition reaction at a potential $\phi^{ox}$ more negative than the OER potential $\phi^0(\mathrm{H_2O/O_2})$ (Figure 1b,d) or at least one reductive decomposition reaction at a potential $\phi^{re}$ more positive than the HER potential $\phi^0(\mathrm{H^+/H_2})$ (Figure 1c,d). Notably, a semiconductor exhibiting a bandgap not straddling the HER and OER potentials could still be stable against photocorrosion, even if the relative positions between the decomposition redox potentials and the water redox potentials are similar, as shown in Figure 1b-d. For instance, if the reductive decomposition potential $\phi^{re}$ is more positive than the HER potential $\phi^0(\mathrm{H^+/H_2})$ but less positive than the CB potential $\phi^{\mathrm{CB}}$, photogenerated electrons prefer to stay at the CB edge instead of driving self-decomposition before recombination.

## III. Results and Discussion

**Validating the theoretical predictions.** Now, we can assess the stability of semiconductors under solar water-splitting conditions using the criterion shown in Figure 1 by determining the energy positions of the band edges and the redox decomposition potentials with respect to the water redox potential for semiconductors. We examined 202 inorganic materials known to either catalyze overall water splitting or reduce or oxidize water in the presence of an external sacrificial redox agent[13-15]. We organized these 202 materials in Figure 2 from left to right in descending order of energy of their VB edge potentials (at pH = 0). The orange color marks the photocorrosion-resistant materials, and the blue color indicates the remaining unstable materials. Details of all the predicted photodecomposition lowest reductive and highest oxidative potentials and reactions are given in Table S1 in the Supplementary Materials. We see that all semiconductors that are stable against photocorrosion are, as expected, oxides or halides with low VB energies.

To test these theoretical predictions, we performed detailed comparisons with experimental measurements for widely studied or recently reported semiconductor photocatalysts, such as $TiO_2$, $Cu_2O$, $BiVO_4$, $Ag_3PO_4$, β-$Ge_3N_4$ and BiOX (X = Cl, Br, and I).

Since the demonstration of PEC water splitting over $TiO_2$ by Fujishima and Honda[16], $TiO_2$ has been the most widely used oxide for photocatalytic applications owing to its low cost and long-term stability in aqueous solution[1-3, 17]. Figure 3a shows that the predicted reductive decomposition potential $\phi^{re}$ and oxidative decomposition potential $\phi^{ox}$ straddle the HER potential $\phi^0(\mathrm{H^+/H_2})$ and OER potential $\phi^0(\mathrm{H_2O/O_2})$, indicating that $TiO_2$ is thermodynamically stable under water-splitting conditions, which is in excellent agreement with experiments. As an active photocatalyst under visible-light illumination, $Cu_2O$ has recently attracted attention[12], even though $Cu_2O$ suffers from poor stability in aqueous solution. We predict that, as shown in Figure 3a, both the reductive and oxidative reactions decompose $Cu_2O$ as a result of the more positive reductive decomposition potential $\phi^{re}$ than the HER potential $\phi^0(\mathrm{H^+/H_2})$ and the much more negative oxidative decomposition potential $\phi^{ox}$ than the OER potential $\phi^0(\mathrm{H_2O/O_2})$, in agreement with the previous experimental report[12]. Here, a slight difference was observed in which our predicted oxidative decomposition potential $\phi^{ox}$ was higher than that determined in an earlier study[12], where both redox potentials lied within the bandgap. In recent years, the visible-light-active $BiVO_4$ semiconductor has also been widely studied as a promising photoanode in water-splitting PEC cells[18]. Two opposite opinions have been reported on the stability of $BiVO_4$ against photocorrosion. Some experiments presented evidence suggesting the material is stable in PEC water splitting[19], whereas experimental evidence[20] has also been represented suggesting degradation occurs, as the photocurrent generated from the bare $BiVO_4$ film decreases significantly within a few minutes under illumination, ascribing to severe anodic photocorrosion occurring on the $BiVO_4$ surface. The observed remarkable reduction of the V/Bi ratio on the $BiVO_4$ surface under illumination was another supporting evidence for the photodecomposition of $BiVO_4$[20]. Here, our prediction suggests $BiVO_4$ is resistant to photocorrosion at pH = 0, as the oxidative decomposition potential ($\phi^{ox} = 1.36$ V *vs* NHE) is more positive than the HER potential $\phi^0(\mathrm{H_2O/O_2})$, and the potential of the CB edge is more positive than the reductive potential ($\phi^{re} = 0.05$ V *vs*

NHE); however, the reductive decomposition potential is more positive than the HER potential $\phi^0(\mathrm{H^+/H_2})$, implying that $BiVO_4$ may become unstable as increasing the pH value. $Ag_3PO_4$ having a 2.4 eV bandgap that absorbs visible light to oxidize water into oxygen[21, 22] has been regarded as chemically stable in aqueous solution[22]. However, experiments[21] have also reported that $Ag^+$ transforms into Ag during the active photocatalytic process in the absence of an electron acceptor, resulting in black metallic Ag particles accumulating on the $Ag_3PO_4$ surface, which suppresses further photocatalytic activity by preventing light absorption and thus indicates photodecomposition of $Ag_3PO_4$. Here, we predict that $Ag_3PO_4$ is stable against photocorrosion, as the CB edge potential is lower than both the reductive decomposition potential $\phi^{re}$ and the HER potential $\phi^0(\mathrm{H^+/H_2})$, and thus, the photogenerated electrons prefer to stay at the CB edge instead of driving HER or reductive photodecomposition; however, the lowest reduction decomposition potential ($\phi^{re} = 0.19$ V *vs* NHE) of $Ag_3PO$ is more positive than the HER potential $\phi^0(\mathrm{H^+/H_2})$. Note that, at a high pH value, $Ag_3PO_4$ becomes unstable because of a slightly more negative reductive decomposition potential (at pH = 0) crossing the energy of the CB edge, as shown in Figure S1e in the Supplementary Materials, which explains the previous reported self-decomposition of $Ag_3PO_4$ [21].

$RuO_2$-loaded β-$Ge_3N_4$ was the first successful example of a non-oxide photocatalyst for overall water splitting[23]. Unfortunately, we found that β-$Ge_3N_4$ was unstable against photocorrosion. Figure 3a shows that the photodecomposition oxidative potential of β-$Ge_3N_4$ is much more negative than the OER potential, indicating that photogenerated holes cause the self-decomposition of β-$Ge_3N_4$ with an oxidation reaction of $Ge_3N_4 + 6H_2O \rightarrow 6H_2 + 3GeO_2 + 2N_2$. The photodecomposition of β-$Ge_3N_4$ was experimentally confirmed by the observation of a relatively high amount of $N_2$ evolution during water splitting[24]. Bismuth oxyhalides, BiOX (where X = Cl, Br or I), have recently been studied[25] as novel non-oxide photocatalysts. A very recent experiment reported that BiOBr exhibited excellent stability in aqueous solution, but both BiOI and BiOCl possessed poor stabilities[26]. These experimental results are consistent with our predictions. BiOBr is photocorrosion resistant, as the two photodecomposition redox potentials ($\phi^{re} = -0.18\,\mathrm{V}, \phi^{ox} = 1.32\,\mathrm{V}$ *vs* NHE) of BiOBr straddle the HER and OER potentials. However, BiOI is susceptible to oxidative photodecomposition, as indicated by the more negative oxidation potential ($\phi^{ox} = 0.96$ V) than the OER potential ($\phi^0(H_2O/O_2) = 1.23$ V *vs* NHE), and BiOCl is subject to reductive photodecomposition because of the slightly more positive reductive potential of $\phi^{re} = 0.04$ V *vs* NHE.

**Photocorrosion-induced limit for the VB potentials of photocatalysts.** Figure 2 indicates that the semiconductors near the left side of the plot are all unstable (marked in blue), whereas most of the stable semiconductors (marked in orange) reside on the right side of the plot. Surprisingly, we observed a line at 2.48 V *vs* NHE in the VB edge potentials, above which no semiconductor photocatalysts were stable against photocorrosion. The emergence of such a limiting line for the photocatalyst VB potential indicates that the photocorrosion sets a limit for the highest possible VB edge potential for a stable semiconductor photocatalyst. We name this limiting line the photocatalyst-limiting line (or PLL, $\phi^{PLL} = 2.48$ V *vs* NHE). Although semiconductors with VB edge potentials below this PLL are not necessarily stable against photocorrosion, a VB edge potential below the PLL is required for a semiconductor photocatalyst to exhibit photocorrosion resistance. The overpotential between the PLL ($\phi^{PLL} = 2.48$ V) and the OER potential ($\phi^0(H_2O/O_2) =$ 1.23 V *vs* NHE) poses a major challenge for the development of high-performance photocatalyst materials. Because the potential of the CB edge must be more negative than the HER potential $\phi^0(H^+/H_2)$, the limit for photocorrosion-resistant photocatalysts with the highest VB potential at 2.48 V excluded PEC photocatalytic materials with bandgaps less than 2.48 eV.

As shown in Figure 2, photocorrosion-resistant semiconductors always possess wide bandgaps and therefore suffer from poor PEC activity under visible light. Several strategies have emerged to overcome this limitation, such as incorporating nitrogen into the metal-oxide lattice to form oxynitrides[27, 28], which decreases the bandgap by raising the VB, as the N *2p* orbital is higher in energy than the O *2p* orbital. Figure 3b shows that the energy positions of the VB edges of all the examined oxynitrides are raised remarkably relative to the VB edges of their oxide counterparts (even passing over the PLL), whereas their CB edges show little change in energy. However, all these oxynitrides are predicted to be unstable, consistent with the manifestation of the PLL, but their oxide counterparts are stable against photocorrosion. For example, the oxidative decomposition potential of TaON is $\phi^{ox} = 0.06$ V *vs* NHE for the oxidative reaction of $2TaON + 3H_2O \rightarrow 3H_2 + N_2 + Ta_2O_5$, indicating that TaON will decompose back to the more stable oxide $Ta_2O_5$ under water-splitting conditions. This prediction was confirmed by experimental observations[27, 28], where a low level of $N_2$ evolution during solar water splitting was observed over this photoelectrode in the initial stages, which was attributed to the partial decomposition of the oxynitride driven by photogenerated holes. In such a self-decomposition progress, a $Ta_2O_5$ thin film built up on top of TaON to serve as an oxide protection layer to suppress the further decomposition of TaON, which explains the experimentally observed ceasing of $N_2$ evolution after the initial stages. Because photogenerated holes are difficult to extract across a newly formed oxide protection layer to drive OER, a low quantum yield (5-6%) for solar water splitting[28] is thus expected using an oxynitrides as a photocatalyst.

The introduction of defects is another attractive method to raise the VB of photocorrosion-resistant oxides. For instance, introducing disorder through hydrogenation in the surface layers of nanophase $TiO_2$ has been examined to extend the absorption edge into the visible-light region[29, 30]. The modified $TiO_2$ (named black $TiO_2$) exhibited activity under visible light. However, experiments also reported that the $H_2$ production rate dropped by two orders of magnitude when black $TiO_2$ was illuminated with visible and infrared light, with incident light of wavelengths shorter than 400 nm filtered out[29, 31]. This observation may be due to the coating of black $TiO_2$ by a layer of wide-bandgap white $TiO_2$, which functions as an oxide protection layer to suppress further degradation but also block photogenerated carriers to drive HER and OER. This speculation is supported by the experimental observation[29] of identical Ti 2p XPS spectra for both white and black $TiO_2$. The coating of the white $TiO_2$ layer could stem from the decomposition of black $TiO_2$, which was revealed to have a nanocrystalline $TiO_2$ core and a highly disordered hydrogen-doped $TiO_2$ shell (ca. 1 nm thick) [31].

We conclude that, although we can modify a photocorrosion-resistant metal oxide to raise the VB edge to extend the bandgap into the visible-light region, the modified metal oxide typically becomes unstable because of the VB edge approaching the PLL, implying that the PLL ($\phi^{PLL} = 2.48$ V *vs* NHE) is universal for PEC semiconducting photocatalysts. Notice that, in practical PEC cells, catalysts are additionally decorated on semiconductor surfaces to promote the OER and HER reactions by lowering their respective overpotentials. These catalysts may slow down the photodecomposition rate of the decorated-semiconductor but will not fully protect it since they absorb sunlight and thus cannot fully cover the surface of semiconductor photocatalyst. Here, we can safely neglect catalyst effect on stability [8-10, 32].

**Revisiting the upper-bound efficiency of single-bandgap PEC water splitting.** The absorbed photon flus $J_g$ (photons $s^{-1}m^{-2}$) is an important quantity in solar efficiency calculations and is defined by $J_g = \int_{E_g}^{\infty} J_{\hbar\upsilon}\alpha_{\hbar\upsilon}d\hbar\upsilon$, here $J_{\hbar\upsilon}$ is the AM1.5 global solar spectrum as shown in Figure 4A. To evaluate the theoretical limiting efficiency, here, we take a common assumption that complete absorption ($\alpha_{\hbar\upsilon}$=1) of all photons above the bandgap of the semiconductor [33, 34]**.** Figure 4B shows the maximum harvested solar energy percentage $P = J_g/\int_0^{\infty} J_{\hbar\upsilon}d\hbar\upsilon$ for a semiconductor with bandgap $E_g$. A semiconductor with a 2.48 eV bandgap can only absorb a small part of the solar spectrum ($E_{\mathrm{phtoton}} \geq 2.48$ eV), accounting for ~23% of solar radiation, as shown in Figure 4B, and possessing a maximum achievable efficiency of 13% for a single-bandgap photovoltaic cell according to the Shockley-Queisser limit[35]. However, for solar water-splitting applications, the maximum achievable STH efficiency is further reduced to 7.8% [33, 34] as shown in Figure 4 (or 7.1% for a realistic single junction PEC cell[34]) because only 1.23 eV, rather than 2.48 eV, per absorbed photon is converted to chemical energy through driving the overall water-splitting reaction. Therefore, the required PEC stability remarkably decreases the STH limiting efficiency of single-bandgap PEC water splitting from the previously thought value of 30.7% to 7.8%, which is even lower than the 10% efficiency required for commercial applications[1, 3]. This finding emphasizes the difficulty of finding a stable semiconducting photocatalyst with a high PEC efficiency even after an extensive search over half a century for materials that simultaneously meet the requirements of both favorable energy band positions and photocorrosion resistance. In the above discussion, we ignore the effect of OER and HER overpotentials, which will further lower the predicted 7.8% upper bound efficiency of PEC water splitting.

**The physics underlying the photocorrosion-induced limit of the VB potentials for photocatalysts.** To uncover the underlying physics, we attempt to derive the relation between the VB potential and free-energy for semiconducting photocatalysts considering the photocorrosion resistance of a semiconducting photocatalyst is highly coupled to the Gibbs free energy of formation. A semiconducting photocatalyst that possesses a more negative Gibbs free energy of formation requires more free energy supplied from photogenerated carriers to drive the decomposition redox reactions and hence is more likely stable against photocorrosion. For example, the Gibbs free energy of formation of $\mathrm{M_aX_b}$ can be written as $\Delta_f G(\mathrm{M_aX_b}) = \Delta_f H(\mathrm{M_aX_b}) - TS$ (*T* is the temperature; *S* is the entropy), and the enthalpy of formation is defined as $\Delta_f H(\mathrm{M_aX_b}) = E_{tot}(\mathrm{M_aX_b}) - a\mu_\mathrm{M} - b\mu_\mathrm{X}$ ($\mu_i$ is the chemical potential of element *i*; $E_\mathrm{tot}(\mathrm{M_aX_b})$ is the total energy of $\mathrm{M_aX_b}$). The enthalpy of formation $\Delta_f H(\mathrm{M_aX_b})$ is highly correlated to the VB energy in which a lower VB energy results in a more negative $\Delta_f H(\mathrm{M_aX_b})$ for a semiconductor. Because the VB edge states are generally the anion atom centered bonding states, a lower VB indicates either a stronger covalent bond or a larger Coulomb binding between anions and cations, which in turn results in a more negative total energy $E_{tot}(\mathrm{M_aX_b})$ (and $\Delta_f H$) for the semiconductor. Furthermore, combining the $E_\mathrm{VB}$ *vs* $E_g$ relationship proposed by Butler and Giney[36] and the $E_g$ *vs* $\Delta_f H$ relationship proposed by Portier *et al.*[37], we can obtain a quantitative relationship between $E_\mathrm{VB}$ and $\Delta_f H$ for a semiconductor $\mathrm{M_aX_b}$ at pH = 0:

$$E_\mathrm{VB} = 0.619\,\chi(\mathrm{M_aX_b}) + 0.5\,A_c\,exp\left(-2.95\times10^{-5}\cdot\Delta_f H/n_e\right) - 1.942 \qquad (1)$$

where $\chi(\mathrm{M_aX_b}) = (\chi_M^a \cdot \chi_X^b)^{1/(a+b)}$, $\chi_M^a$ and $\chi_X^b$ are the absolute electronegativities of atoms M and X, respectively, $n_e$ is the number of electrons involved in the reaction, $A_c$ is a property of the cation, and $E_\mathrm{VB}$ and $\Delta_f H$ are in eV (see the Supplementary Materials for more details).

Based on the deduced $E_\mathrm{VB}$ *vs* $\Delta_f H$ relationship, we could verify the photocorrosion-induced limit of the VB potentials by artificially raising the VB edge potentials of all the stable semiconducting photocatalysts to 2.48 V *vs* NHE and adjusting $\Delta_f H$ according to Eq. (1). In this case, we found that all the semiconductors again become unstable. For example, the VB edge potential of $TiO_2$ was at 3.25 V *vs* NHE, and the Gibbs free energy was -888.8 KJ/mol with predicted reductive and oxidative photodecomposition potentials of $\phi^{\mathrm{re}} = -0.53$ V and $\phi^{\mathrm{ox}} = 1.30$ V, respectively (corresponding reactions are given in Table S1). If we artificially raise the VB edge potential from 3.25 to 2.48 V *vs* NHE, the Gibbs free energy changes from -888.8 to -801.50 KJ/mol according to Eq. (1), and $TiO_2$ becomes unstable, as the modified reductive potential ($\phi^{\mathrm{re}} = 0.78$ V) becomes more positive than the HER potential. If we upshift the VB edge potential of $Bi_2WO_6$ from 3.27 to 2.48 V, the Gibbs free energy changes from -1708.34 to -1439.69 KJ/mol, and the reductive and oxidative photodecomposition potentials change from $\phi^{\mathrm{re}} = -0.25$ to 0.07 V and from $\phi^{\mathrm{ox}} = 1.28$ to 0.28 V, respectively. Subsequently, $Bi_2WO_6$ again becomes unstable upon upshifting the VB potential above the PLL. Thereby, we have illustrated that the photocorrosion-induced limit of the VB potentials for photocatalysts has a deep root in underlying physics.

## IV. Conclusions.

We revealed that the thermodynamic stability is strongly coupled to the energy position of the VB edge for semiconducting PEC photocatalysts, setting limits for the highest possible VB edge potential at 2.48 V *vs* NHE (named the PLL) and for the minimum possible bandgap at 2.48 eV. These limits remarkably reduce the STH limiting efficiency for single-bandgap PEC water splitting from the commonly believed 30.7% to approximately 8%, even lower than the 10% efficiency required for commercial applications[1-3]. Although this conclusion was drawn from data at pH = 0, the limits are also applicable in a finite pH range (see discussion on pH dependence in the Supplementary Materials). Because the predicted limiting STH efficiency is too low, future efforts toward attaining highly efficient PEC solar water splitting must shift from searching for PEC photocatalysts to strategies that decouple the thermodynamic stability and PEC efficiency. For example, metal-oxide protection layers have been used to stabilize narrow-bandgap (PEC-active) semiconducting photocatalysts, such as $Cu_2O$[12], Si[6, 38, 39], CdTe[39], and III-V[7] semiconductors as well as some modified oxides. For this purpose, the focus should shift toward interface engineering for the simultaneous optimization of the built-in field, interface quality, and carrier extraction to maximize the photovoltage of an oxide-protected water-splitting photocatalyst[32, 38, 40, 41]. Furthermore, a dual-bandgap Z-scheme water-splitting system using two different semiconducting photocatalysts could also overcome the predicted 8% limiting STH efficiency if the photoanode is resistant to photoanodic decomposition and if the photocathode is resistant to photocathodic decomposition.

## Acknowledgments

We thank S. Chen for helpful discussion and carefully reading the manuscript. J.W.L thanks A. Zunger for helpful discussions. J.W.L. was supported by the National Natural Science Foundation of China (NSFC) under Grant No. 61474116 and the National Young 1000 Talents Plan. L.J.G was supported by NSFC under Grant No. 11404074. S.H.W. was supported by the NSFC under Grant No. U1530401 and 51672023 and the National Key Research and Development Program of China under Grant No. 2016YFB0700700.

**References:**

1. Gratzel, M., *Photoelectrochemical cells.* Nature, 2001. **414**(6861): p. 338-344.
2. Walter, M.G., et al., *Solar Water Splitting Cells.* Chemical Reviews, 2010. **110**(11): p. 6446-6473.
3. Osterloh, F.E. and B.A. Parkinson, *Recent developments in solar water-splitting photocatalysis.* MRS Bulletin, 2011. **36**(01): p. 17-22.
4. Hisatomi, T., J. Kubota, and K. Domen, *Recent advances in semiconductors for photocatalytic and photoelectrochemical water splitting.* Chem. Soc. Rev., 2014. **43**(22): p. 7520-7535.
5. Ager, J.W., et al., *Experimental demonstrations of spontaneous, solar-driven photoelectrochemical water splitting.* Energy & Environmental Science, 2015. **8**(10): p. 2811-2824.
6. Kenney, M.J., et al., *High-Performance Silicon Photoanodes Passivated with Ultrathin Nickel Films for Water Oxidation.* Science, 2013. **342**(6160): p. 836-840.
7. Gu, J., et al., *Water reduction by a p-GaInP2 photoelectrode stabilized by an amorphous TiO2 coating and a molecular cobalt catalyst.* Nature Materials, 2015. **15**(4): p. 456-460.
8. Castelli, I.E., et al., *Computational screening of perovskite metal oxides for optimal solar light capture.* Energy & Environmental Science, 2012. **5**(2): p. 5814-5819.
9. Wu, Y., et al., *First principles high throughput screening of oxynitrides for water-splitting photocatalysts.* Energy & Environmental Science, 2013. **6**(1): p. 157-168.
10. Chen, S. and L.-W. Wang, *Thermodynamic Oxidation and Reduction Potentials of Photocatalytic Semiconductors in Aqueous Solution.* Chemistry of Materials, 2012. **24**(18): p. 3659-3666.
11. Gerischer, H., *On the stability of semiconductor electrodes against photodecomposition.* Journal of Electroanalytical Chemistry and Interfacial Electrochemistry, 1977. **82**(1-2): p. 133-143.
12. Paracchino, A., et al., *Highly active oxide photocathode for photoelectrochemical water reduction.* Nat Mater, 2011. **10**(6): p. 456-461.
13. Xu, Y. and M.A.A. Schoonen, *The absolute energy positions of conduction and valence bands of selected semiconducting minerals.* American Mineralogist, 2000. **85**(3-4): p. 543-556.
14. Kudo, A. and Y. Miseki, *Heterogeneous photocatalyst materials for water splitting.* Chem Soc Rev, 2009. **38**(1): p. 253-78.
15. Chen, X., et al., *Semiconductor-based photocatalytic hydrogen generation.* Chem Rev, 2010. **110**(11): p. 6503-70.
16. Fujishima, A. and K. Honda, *Electrochemical Photolysis of Water at a Semiconductor Electrode.* Nature, 1972. **238**(5358): p. 37-38.
17. Kazuhito, H., I. Hiroshi, and F. Akira, *TiO 2 Photocatalysis: A Historical Overview and Future Prospects.* Japanese Journal of Applied Physics, 2005. **44**(12R): p. 8269.
18. Park, Y., K.J. McDonald, and K.S. Choi, *Progress in bismuth vanadate photoanodes for use in solar water oxidation.* Chem Soc Rev, 2013. **42**(6): p. 2321-37.
19. Kudo, A., K. Omori, and H. Kato, *A Novel Aqueous Process for Preparation of Crystal Form-Controlled and Highly Crystalline BiVO4Powder from Layered Vanadates at Room Temperature and Its Photocatalytic and Photophysical Properties.* Journal of the American Chemical Society, 1999. **121**(49): p. 11459-11467.

20. Seabold, J.A. and K.S. Choi, *Efficient and stable photo-oxidation of water by a bismuth vanadate photoanode coupled with an iron oxyhydroxide oxygen evolution catalyst.* J Am Chem Soc, 2012. **134**(4): p. 2186-92.
21. Yi, Z., et al., *An orthophosphate semiconductor with photooxidation properties under visible-light irradiation.* Nature Materials, 2010. **9**(7): p. 559-564.
22. Chen, X., Y. Dai, and X. Wang, *Methods and mechanism for improvement of photocatalytic activity and stability of Ag3PO4: A review.* Journal of Alloys and Compounds, 2015. **649**(Complete): p. 910-932.
23. Sato, J., et al., *RuO2-Loaded β-Ge3N4 as a Non-Oxide Photocatalyst for Overall Water Splitting.* Journal of the American Chemical Society, 2005. **127**(12): p. 4150-4151.
24. Lee, Y., et al., *Effect of High-Pressure Ammonia Treatment on the Activity of Ge3N4 Photocatalyst for Overall Water Splitting.* The Journal of Physical Chemistry B, 2006. **110**(35): p. 17563-17569.
25. Ye, L., et al., *Recent advances in BiOX (X = Cl, Br and I) photocatalysts: synthesis, modification, facet effects and mechanisms.* Environmental Science: Nano, 2014. **1**(2): p. 90.
26. Bhachu, D.S., et al., *Bismuth oxyhalides: synthesis, structure and photoelectrochemical activity.* Chem. Sci., 2016. **7**(8): p. 4832-4841.
27. Maeda, K. and K. Domen, *Oxynitride materials for solar water splitting.* MRS Bulletin, 2011. **36**(01): p. 25-31.
28. Takata, T., C. Pan, and K. Domen, *Recent progress in oxynitride photocatalysts for visible-light-driven water splitting.* Science and Technology of Advanced Materials, 2015. **16**(3): p. 033506.
29. Chen, X., et al., *Increasing solar absorption for photocatalysis with black hydrogenated titanium dioxide nanocrystals.* Science, 2011. **331**(6018): p. 746-50.
30. Deng, H.-X., et al., *Effect of hydrogen passivation on the electronic structure of ionic semiconductor nanostructures.* Physical Review B, 2012. **85**(19): p. 195328.
31. Hu, Y.H., *A highly efficient photocatalyst--hydrogenated black TiO2 for the photocatalytic splitting of water.* Angew Chem Int Ed Engl, 2012. **51**(50): p. 12410-2.
32. Pham, T.A., Y. Ping, and G. Galli, *Modelling heterogeneous interfaces for solar water splitting.* Nature Materials, 2017. **16**(4): p. 401-408.
33. Bolton, J.R., S.J. Strickler, and J.S. Connolly, *Limiting and realizable efficiencies of solar photolysis of water.* Nature, 1985. **316**(6028): p. 495-500.
34. Fountaine, K.T., H.J. Lewerenz, and H.A. Atwater, *Efficiency limits for photoelectrochemical water-splitting.* Nature Communications, 2016. **7**: p. 13706.
35. Shockley, W. and H.J. Queisser, *Detailed Balance Limit of Efficiency of p‐n Junction Solar Cells.* Journal of Applied Physics, 1961. **32**(3): p. 510-519.
36. Butler, M.A., *Prediction of Flatband Potentials at Semiconductor-Electrolyte Interfaces from Atomic Electronegativities.* Journal of The Electrochemical Society, 1978. **125**(2): p. 228.
37. Portier, J., et al., *Relationships between optical band gap and thermodynamic properties of binary oxides.* International Journal of Inorganic Materials, 2001. **3**(7): p. 1091-1094.
38. Scheuermann, A.G., et al., *Design principles for maximizing photovoltage in metal-oxide-protected water-splitting photoanodes.* Nature Materials, 2015. **15**(1): p. 99-105.

39. Sun, K., et al., *Stable solar-driven oxidation of water by semiconducting photoanodes protected by transparent catalytic nickel oxide films.* Proceedings of the National Academy of Sciences, 2015. **112**(12): p. 3612-3617.
40. Liu, R., et al., *Enhanced photoelectrochemical water-splitting performance of semiconductors by surface passivation layers.* Energy & Environmental Science, 2014. **7**(8): p. 2504-2517.
41. Yang, Y., et al., *Semiconductor interfacial carrier dynamics via photoinduced electric fields.* Science, 2015. **350**(6264): p. 1061.

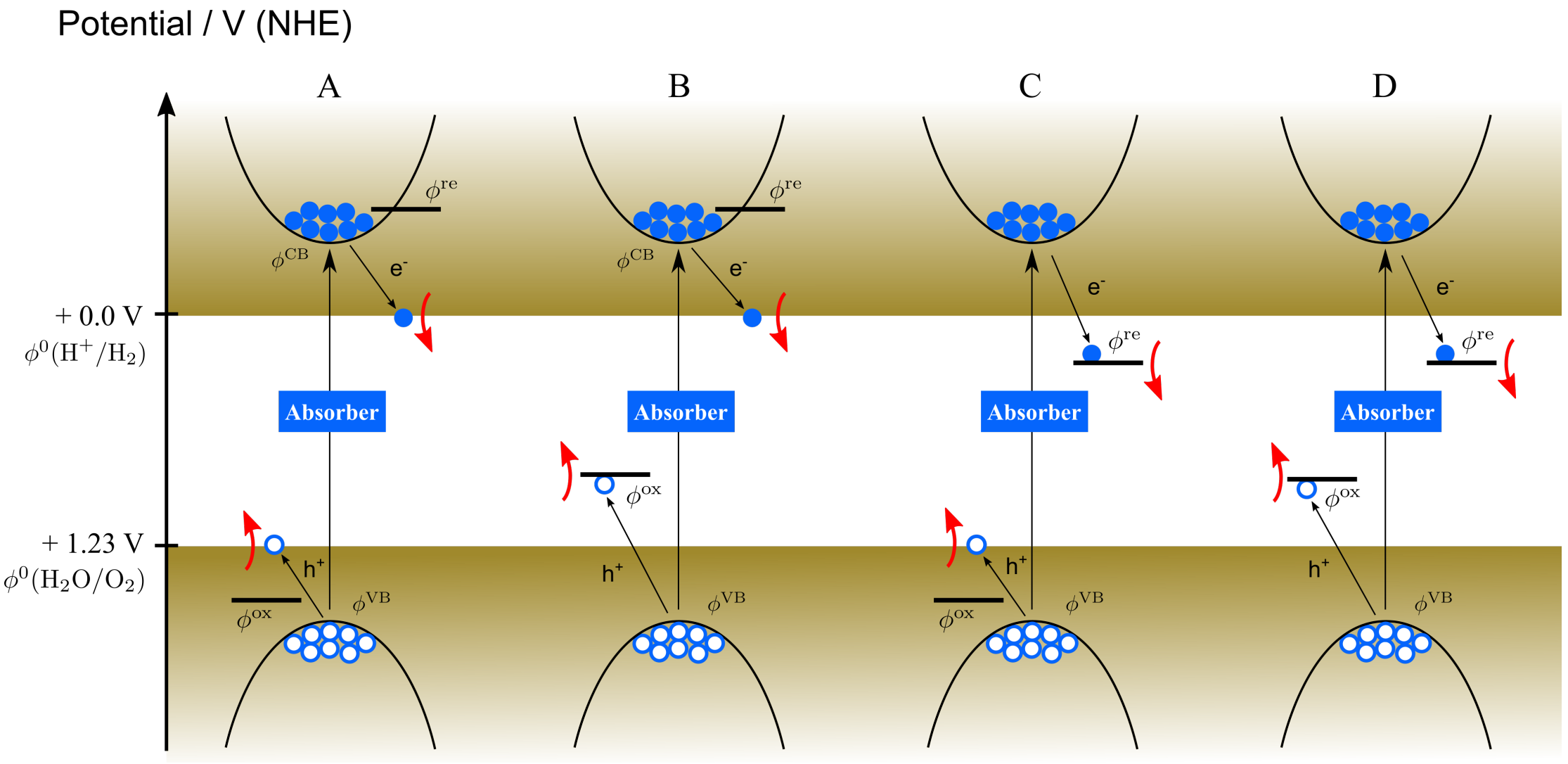

**Figure 1. Classification of the semiconductor decomposition redox potentials with respect to the water redox potentials.** For optimized PEC water splitting, a semiconductor should possess a bandgap straddling the HER and OER potentials. The relative energy positions between the band-edge energy, the water redox potentials and the semiconductor decomposition redox potentials characterize the stability of the semiconductor against photocorrosion. (**A**) The semiconductor is stable against photocorrosion because of a more negative reductive photodecomposition potential $\phi^{re}$ than the HER potential $\phi^0(H^+/H_2)$ and a more positive oxidative photodecomposition potential $\phi^{ox}$ than the OER potential $\phi^0(H_2O/O_2)$. (B) The semiconductor is susceptible to anodic photodecomposition because of the less negative reductive photodecomposition potential $\phi^{re}$ than the HER potential $\phi^0(H^+/H_2)$. (C) The semiconductor is susceptible to reductive photodecomposition because of the less positive oxidative photodecomposition potential $\phi^{ox}$ than the OER potential $\phi^0(H_2O/O_2)$. (D) The semiconductor is susceptible to both reductive and oxidative photodecompositions.

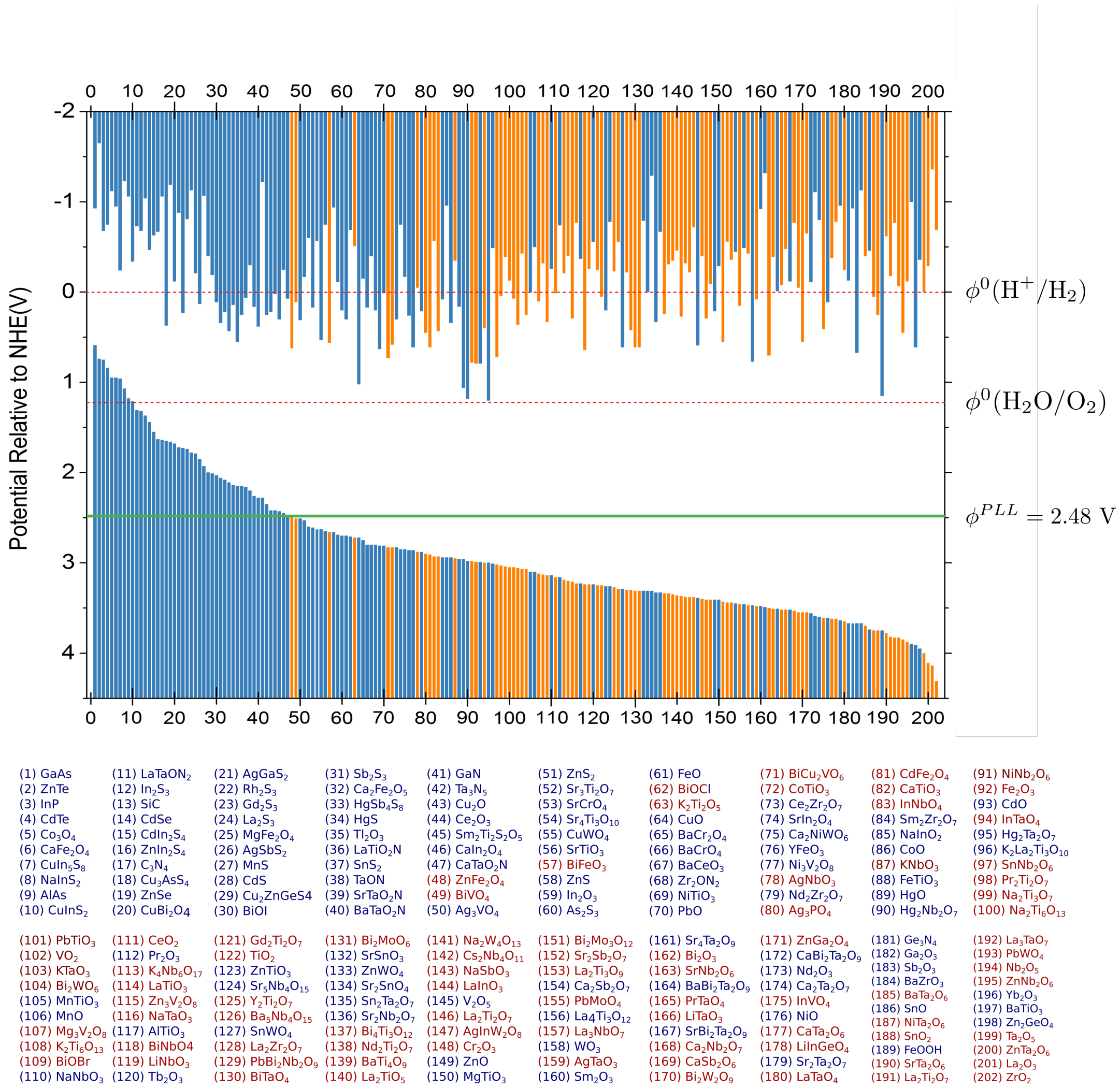

**Figure 2. Band alignments and stabilities of 202 semiconductors under water-splitting conditions.** The stabilities of 202 inorganic materials known to either catalyze overall water splitting or reduce or oxidize water in the presence of an external sacrificial redox agent under water-splitting conditions are evaluated. Photocorrosion-resistant semiconductor photocatalysts are marked in the orange color, and the remaining unstable semiconductor photocatalysts are indicated by the blue color. These 202 materials are organized from left to right in descending order of energy of their VB edge potentials (at pH = 0). The emergence of a photocatalyst line in the VB edge potentials (analogous to the tree line on a mountain) is evident. Above this line, no semiconductor photocatalyst is stable against photocorrosion.

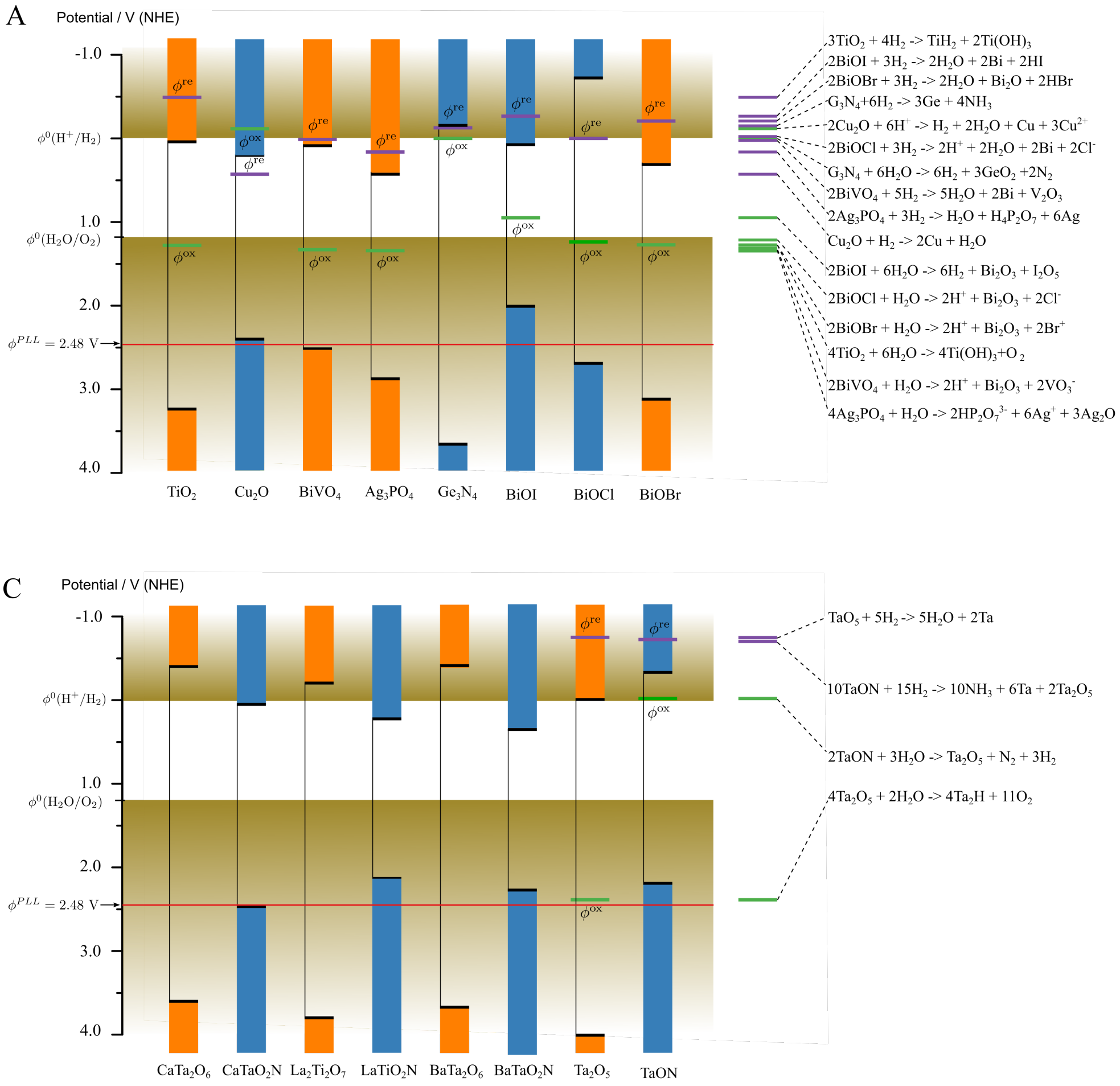

**Figure 3. Stability of various semiconductors and oxynitrides under solar water-splitting conditions. (A)** The energy positions (versus NHE) of the CB and VB edges of various selected semiconductors, including anatase $TiO_2$, $Cu_2O$, $BiVO_4$, β-$Ge_3N_4$, and BiOX (X = I, Cl, or Br) at pH = 0, and the photodecomposition redox potentials under solar water-splitting conditions. Photocorrosion-resistant semiconductor photocatalysts are marked in the orange color, and the remaining unstable semiconductor photocatalysts are indicated by the blue color. The corresponding photodecomposition redox reactions of these semiconductors are listed on the right. (B) The energy positions of the CB and VB edges of various oxynitrides and their oxide counterparts.

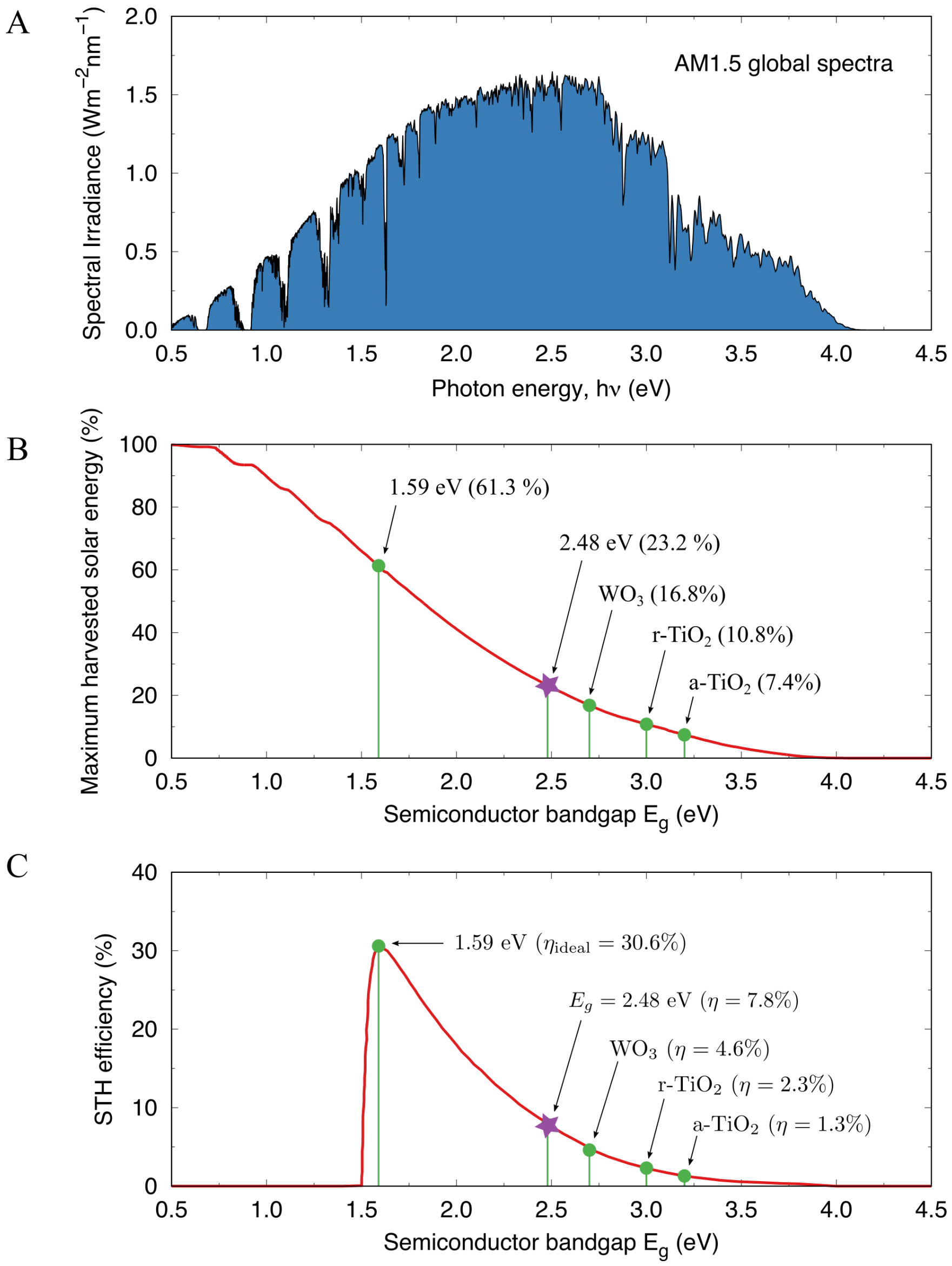

**Figure 4. Limiting STH efficiency versus the semiconductor bandgap (*Eg*) for single-absorber PEC cells. (A)** Solar spectral distribution based on AM1.5 global data. **(B)** Maximum harvested solar energy percentage of semiconductor absorbers versus the semiconductor bandgap *Eg*. The values corresponding to the optimal bandgap (1.59 eV), the photocorrosion-limited smallest bandgap (2.48 eV), $WO_3$, rutile $TiO_2$, and anatase $TiO_2$ are also indicated. **(C)** The maximum STH efficiencies of the single-absorber PEC cells for the ideal case without considering stability ($Eg$ = 1.59 eV, $\eta$ = 30.6%), for the ideal case considering the photocorrosion restriction ($Eg$ = 2.48 eV, $\eta$ = 7.8%), and for the cases of $WO_3$, rutile $TiO_2$, and anatase $TiO_2$ cells.